\documentclass[lettersize,journal]{IEEEtran}

\usepackage{amsmath,amsfonts}
\usepackage{algorithmic}
\usepackage{array}
\usepackage{textcomp}
\usepackage{stfloats}
\usepackage{url}
\usepackage{verbatim}
\usepackage{xcolor}
\usepackage{graphicx}
\usepackage{array}
\usepackage{tabularx}
\usepackage{amssymb}
\usepackage{multicol}
\usepackage{caption}

\usepackage{subcaption}
\allowdisplaybreaks
\def\BibTeX{{\rm B\kern-.05em{\sc i\kern-.025em b}\kern-.08em
		T\kern-.1667em\lower.7ex\hbox{E}\kern-.125emX}}
\usepackage{balance}

\begin{document}
\title{Collision Resolution in RFID Systems Using Antenna Arrays and Mix Source Separation
		}
	\author{Mohamed Siala, and Noura Sellami	
    \thanks{Mohamed Siala is with MEDIATRON Lab., SUP'COM, University of Carthage, Tunisia (email: mohamed.siala@supcom.tn). \\ Noura Sellami is with  LETI Lab., ENIS, University of Sfax, Tunisia (email: noura.sellami@enis.tn).}}
\maketitle			
	\begin{abstract}
		In this letter, we propose an efficient mix source separation algorithm for collision resolution in radio frequency identification (RFID) systems equipped with an antenna array at the reader. We first introduce an approach that exploits the zero constant modulus (ZCM) criterion to separate colliding tags through gradient descent, without using pilot symbols. We show that the ZCM characteristic, considered alone, in the design of the objective function can lead to significant ambiguities in the determination of the beamformers used in the recovery of tag messages. To address this limitation, we propose a more sophisticated approach, relying on a hybrid objective function, incorporating a new ambiguity-raising criterion in addition to the ZCM criterion.
			\end{abstract}
\section{Introduction}
Radio frequency identification (RFID) has gained great attention from industry and academia since it plays a crucial role in areas such as supply chain management, security, access control and transportation \cite{ref1}. An RFID system includes three primary components: a tag, a reader and a data collection device. When numerous tags are activated simultaneously, their messages can collide and interfere with each other at the reader. In this case, a retransmission of the tag identities is required. This results not only in a waste of bandwidth, but also in an increase of the overall delay in identifying the objects \cite{ref2}. To address collisions between tag messages, two main approaches can be employed. From a networking perspective, the first strategy makes use of collision avoidance techniques like the tree-splitting algorithm or the Aloha protocol \cite{ref3}. From a signal processing perspective, the second approach focuses on source separation techniques \cite{ref4}.
As most RFID tags lack training symbols, estimating RFID channels is often a challenging task. As a result, mix source separation (MSS) is strongly recommended as a viable approach to overcome tag identification problems \cite{ref5}. In \cite{ref6}, a MSS receiver is introduced for resolving collisions in RFID systems with multiple antennas, utilizing the zero constant modulus (ZCM) criterion. In \cite{ref7}, two anti-collision algorithms are proposed for multiple-antenna RFID systems by combining the ALOHA protocol with MSS. In \cite{ref8}, an RFID collision detection algorithm based on underdetermined mix separation is proposed. 
\\
In this letter, like in \cite{ref6}, we separate the colliding messages of the tags by leveraging the ZCM character of the signals, without using pilot symbols. However, instead of using the algebraic ZCM algorithm (AZCMA), proposed in \cite{ref6}, which requires the resolution of a matrix pencil problem, we apply an objective function minimization, using the gradient descent algorithm. We highlight the potential ambiguities in the determination of source beamformers when the ZCM criterion is used alone and the tag signals are synchronized in frequency with the receiver. These ambiguities lead to spurious beamformers that can enforce the ZCM criterion.  In order to systematically get rid of any ambiguities in the determination of the beamformers, even if perfect or near-perfect frequency synchronization is achieved, we propose a new criterion that is almost never verified by spurious beamformers. We then introduce a new hybrid objective function, mixing the ZCM criterion and the new criterion, and present the corresponding gradient descent algorithm. Simulation results show that the ZCM criterion proposed in \cite{ref6} is not able to separate colliding tag signals, when perfect or almost perfect frequency synchronization is achieved. They also prove the effectiveness of our proposed hybrid objective function in overcoming this limitation.

The remainder of this letter is organized as follows. In Section II, we introduce the system model. In Section III, we recall the AZCMA principle and focus on the ambiguities that it can lead to. In Section IV, we present our new hybrid objective function and its minimization through the gradient descent algorithm. In Section V, we provide simulation results. In Section VI, we conclude the letter. Throughout this letter, vectors and matrices are denoted in bold, in lowercase and uppercase respectively. Moreover, $(.)^*$, $(.)^T$ and $(.)^H$ denote conjugation, transposition, and trans-conjugation, respectively.
\section{System model}

We consider an RFID system operating in the ultrahigh frequency band (UHF) band (860 MHz – 960 MHz), where $d$ tags communicate with a single reader. The tags are equipped with one antenna each, while the reader is equipped with one transmit antenna and $M$ receive antennas. Following the ISO/IEC18000-6 standard \cite{ref9}, the tag responses can be considered as frequency modulated (FM0) signals or as Manchester coding modulated signals. In this letter, we will consider Manchester encoding. Indeed, the two encoding schemes, FM0 and Manchester, become equivalent when one of them is shifted by half the symbol duration with respect to the other, resulting in identical performance regarding tag separation and identification. 
RFID tag responses are organized into data packets of $N$ encoded symbols. Following \cite{ref6}, the $n$-th symbol emanating from the $i$-th tag, $1 \leq i \leq d$, is modeled as
\begin{equation}
s_i[n] = b_i[n] \exp(j \phi_i[n]), \quad 1 \leq n \leq N,
\label{symb}
\end{equation}
where $b_i[n]$ is the binary message of the $i$-th tag and $\phi_i[n]$ is a random phase. If imperfections, due to oscillator drifts and phase fluctuations, are negligible within a received data packet, the random phase $\phi_i[n]$ could be set to a constant $\phi_i$. Complying with the ISO/IEC 18000-6 standard, the $k$-th logical bit of a packet of the $i$-th tag is transmitted through the pair $(b_i[2k], b_i[2k+1])$. Thus, \(N\) is assumed to be an even number. For Manchester coding, the $k$-th logic 1 is sent through $(b_i[2k], b_i[2k+1]) = (1, 0)$, while the $k$-th logic 0 is sent through $(b_i[2k], b_i[2k+1]) = (0, 1)$.

In addition to the tag signals, a strong copy of the reader signal is also present at the receiver \cite{ref6}. The received baseband signal sampled at the symbol rate at the $M$ antennas of the reader at time $n$ is an $M \times 1$ vector $\mathbf{x}[n]$ consisting of a linear combination of the $d$ transmitted tag signals plus the reader signal, perturbed with noise. To achieve spatial decorrelation, the receiving antennas should be separated by at least half a wavelength, which corresponds to approximately 16 cm at 900 MHz in the UHF band. After collecting the $N$ samples corresponding to a data packet, an $M \times N$ data matrix $\mathbf{X} = [\mathbf{x}[1], \mathbf{x}[2], \cdots, \mathbf{x}[N]]$ is created, consisting of $N$ complex vector samples from the $M$ antennas. The multipath delay spread is assumed to be small relative to the symbol duration, and therefore a narrowband channel model is adopted. In most practical cases, the receiver has access to the transmitted reader signal. Hence, following \cite{ref6}, we assume that the reader signal is subtracted from the received data matrix $\mathbf{X}$. Thus, the data matrix can be expressed as
\begin{equation}
\mathbf{X} = \mathbf{A} \mathbf{S} + \mathbf{N}, 
\end{equation}
where $\mathbf{A} = [\mathbf{a}_1, \mathbf{a}_2, \cdots, \mathbf{a}_d]$ is the $M \times d$ channel matrix from the $d$ tags to the $M$ receive antennas, $\mathbf{S} = [\mathbf{s}_1^T, \mathbf{s}_2^T, \cdots, \mathbf{s}_d^T]^T$ is the $d \times N$ matrix stacking the $1 \times N$ signal vectors $\mathbf{s}_i$, for $1 \leq i \leq d$, and $\mathbf{N}$ is an $M \times N$ matrix accounting for time and space decorrelated Gaussian noise samples.

The signal samples in (\ref{symb}) are ZCM, meaning that either $s_i[n] = 0$ or $\lvert s_i[n] \rvert = 1$, for an arbitrary tag $i$, for $1 \leq n \leq N$. In the following, we start by briefly recalling the principle of the AZCMA used in \cite{ref6} and highlighting the ambiguities that can occur in determining the beamformers.
\section{AZCMA and Associated Ambiguities}
Given a received data matrix, $\mathbf{X}$, the aim of the AZCMA is to derive the $M \times 1$ beamformer vectors $\mathbf{w}_i$, for $1 \leq i \leq d$, to recover estimates of the tag signal vectors $\mathbf{s}_i$, as $\hat{\mathbf{s}}_i = \mathbf{w}_i^H \mathbf{X}$, obeying the ZCM property. To enable this source separation, the matrix $\mathbf{A}$ is assumed to be either tall or square, and of full column rank, resulting in an overdetermined or determined system, respectively. In \cite{ref6}, it is shown that the ZCM separation problem is equivalent to finding all solutions to a matrix pencil problem when $N$ is at least equal to $M^3$. In the following, we show that the ZCM nature of the exchanged signals could be at the origin of multiple ambiguities in the determination of the beamformers, $\mathbf{w}_i$, for $1 \leq i \leq d$, of the $d$ tags, whenever imperfections, due to oscillator drifts and phase fluctuations, are negligible. To see this in the easiest possible way, we assume that the noise is negligible and the beamforming is perfectly achieved. We also consider that $\phi_i[n]=\phi_i$, for $1\leq n \leq N$ meaning a perfect synchronisation and no fluctuations in phase, which is a realistic scenario in practice, especially with the technological evolution of tags. Thus, given the modular nature of the metric used by the ZCM, the beamforming reproduces the detected symbols up to a constant phase $\theta_i$. We obtain
\begin{equation}
    \hat{s}_i[n] = \mathbf{w}_i^H \mathbf{x}[n] = e^{j\theta_i}s_i[n]=e^{j\psi_i}b_i[n],
\end{equation}
where $\psi_i = \theta_i + \phi_i$. 

We illustrate the ambiguity problem by considering scenarios involving two or three tags. First, let $\mathbf{w}_{i_1}$ and $\mathbf{w}_{i_2}$ be the beamformer vectors of two arbitrary tags in the system and consider the two alternative beamformers
\begin{equation}
    \mathbf{w}^\pm = e^{j\psi_{i_1}} \mathbf{w}_{i_1} + e^{j(\pm 2\pi/3 + \psi_{i_2})} \mathbf{w}_{i_2}. \label{w+-}
\end{equation}
Then, the corresponding outcomes are given by
\begin{equation}
    (\mathbf{w}^\pm)^H \mathbf{x}[n] = b_{i_1}[n] + e^{\mp j 2\pi/3} b_{i_2}[n]. 
\end{equation}
As shown in Table I, the different outcomes for beamformers $\mathbf{w}^-$ and $\mathbf{w}^+$ have always the ZCM property. As a consequence, these alternative beamformers, which inevitably lead to a mix of the signals of the two considered tags, could potentially be solutions to the AZCMA in addition to the desired beamformers $\mathbf{w}_{i_1}$ and $\mathbf{w}_{i_2}$.

Now, let $\mathbf{w}_{i_1}$, $\mathbf{w}_{i_2}$, and $\mathbf{w}_{i_3}$ be the beamformer vectors of three arbitrary tags, among the $d$ tags in the system, and, again, consider the two new alternative beamformers
\begin{equation}
    \mathbf{w}^\pm = e^{j\psi_{i_1}} \mathbf{w}_{i_1} + e^{j(\pm 2\pi/3 + \psi_{i_2})} \mathbf{w}_{i_2} + e^{j(\mp 2\pi/3 + \psi_{i_3})} \mathbf{w}_{i_3}. \label{w+-3}
\end{equation}
Clearly, the possible outcomes are
\begin{equation}
(\mathbf{w}^\pm)^H \mathbf{x}[n] = b_{i_1}[n] + \exp(\mp j \frac{2\pi}{3}) b_{i_2}[n] + \exp(\pm j \frac{2\pi}{3}) b_{i_3}[n].
\end{equation}
As in Table I, we can easily show that the different outcomes of the beamformers $\mathbf{w}^-$ and $\mathbf{w}^+$ still have the ZCM property and might be additional solutions to the AZCMA, beside $\mathbf{w}_{i_1}$, $\mathbf{w}_{i_2}$, and $\mathbf{w}_{i_3}$. Thus, it becomes clear that any pair or triplet of actual tags’ beamformers can lead to two additional spurious alternative beamformer solutions to the AZCMA.

\begin{table}[h!]
\centering
\caption{Alternative Beamformers’ Outcomes for Two Tags and Their ZCM Properties}
\scriptsize
\resizebox{\columnwidth}{!}{%
\begin{tabular}{|c|c|c|c|c|c|}
\hline
$b_{i_1}[n]$ & $b_{i_2}[n]$ & $(\mathbf{w}^-)^H \mathbf{x}[n]$ & $|(\mathbf{w}^-)^H \mathbf{x}[n]|$ & $(\mathbf{w}^+)^H \mathbf{x}[n]$ & $|(\mathbf{w}^+)^H \mathbf{x}[n]|$ \\
\hline
0 & 0 & 0 & 0 & 0 & 0 \\
\hline
1 & 0 & 1 & 1 & 1 & 1 \\
\hline
0 & 1 & $\exp(j \frac{2\pi}{3})$ & 1 & $\exp(-j \frac{2\pi}{3})$ & 1 \\
\hline
1 & 1 & $\exp(j \frac{\pi}{3})$ & 1 & $\exp(-j \frac{\pi}{3})$ & 1 \\
\hline
\end{tabular}%
}
\end{table}

To get rid of these potential spurious beamformers, a solution is to bring some artificial imperfections like frequency shifts or phase fluctuations at the tags. This solution was implicitly adopted in \cite{ref6}. However, in practice, the system cannot be expected to deliberately exhibit such artificial imperfections. In the following, we propose a new criterion which helps to discern the true beamformers from the spurious ones. We use the gradient descent algorithm in order to minimize a hybrid objective function mixing the ZCM criterion and the new criterion.

\section{Gradient Descent Algorithm Minimizing a New Hybrid Objective Function}
\subsection{New Criterion for Unambiguous Source Separation}

To get rid of any ambiguity in the identification of true beamformers, the approach followed in this letter is to find an additional criterion, that is never or hardly ever verified by spurious beamformers. For the sake of simplicity, we omit in the following the index $i$ which refers to the $i$-th tag. To identify such a criterion, we consider the product
\begin{equation}
\begin{split}
\pi[n] &= \hat{s}[n-1] \hat{s}^*[n] \hat{s}[n+1] \\
&= \mathbf{w}^H \mathbf{x}[n-1] \mathbf{x}^H[n] \mathbf{w} \mathbf{w}^H \mathbf{x}[n+1], \quad 2 \leq n \leq N-1.
\end{split}
\label{new_crit}
\end{equation}
One can easily verify that this product is always null, given that a true beamformer is used and a noiseless reception is assumed. However, if one of the spurious beamformers considered in (\ref{w+-}) or in (\ref{w+-3}) is used, the product $\pi[n]$ can take non-null values with a non-null probability. This is illustrated in Table~\ref{table:ambiguous_beamformer}, where the spurious beamformer, $\mathbf{w}^+$, specified in (\ref{w+-}), is adopted. In Table~\ref{table:ambiguous_beamformer}, only the combinations, among $4 \times 4 = 16$, which lead to at least one non-null product are shown. In the following, we propose to use the gradient descent algorithm to minimize a hybrid objective function mixing the ZCM criterion and the new criterion given in (\ref{new_crit}).

\begin{table*}[ht]
\scriptsize
    \centering
    \caption{NEW CRITERION CHARACTERISTICS WHEN THE SPURIOUS BEAMFORMER OF (\ref{w+-}) IS USED  }
    \label{table:ambiguous_beamformer}
    \begin{tabular}{|c|c|c|c|c|c|c|c|}
    \hline
    $b_{i_1}[n]_{n=2k}^{2k+3}$ & $b_{i_2}[n]_{n=2k}^{2k+3}$ & $\hat{s}[2k]$ & $\hat{s}[2k+1]$ & $\hat{s}[2k+2]$ & $\hat{s}[2k+3]$ & $\pi[2k+1]$ & $\pi[2k+2]$ \\ \hline
    0101 & 0110 & 0 & $e^{j\frac{\pi}{3}}$ & $e^{j\frac{2\pi}{3}}$ & 1 & 0 & $e^{-j\frac{\pi}{3}}$ \\ 
         & 1010 & $e^{j\frac{2\pi}{3}}$ & 1 & $e^{j\frac{2\pi}{3}}$ & 1 & $e^{-j\frac{2\pi}{3}}$ & $e^{-j\frac{2\pi}{3}}$ \\ \hline
    1001 & 0110 & 1 & $e^{j\frac{2\pi}{3}}$ & $e^{j\frac{2\pi}{3}}$ & 1 & 1 & 1 \\ \hline
    0110 & 0101 & 0 & $e^{j\frac{\pi}{3}}$ & 1 & $e^{j\frac{2\pi}{3}}$ & 0 & -1 \\ 
         & 1001 & $e^{j\frac{2\pi}{3}}$ & 1 & 1 & $e^{j\frac{2\pi}{3}}$ & $e^{j\frac{2\pi}{3}}$ & $e^{j\frac{2\pi}{3}}$ \\ 
         & 1010 & $e^{j\frac{2\pi}{3}}$ & 1 & $e^{j\frac{\pi}{3}}$ & 0 & -1 & 0 \\ \hline
    1010 & 0101 & 1 & $e^{j\frac{2\pi}{3}}$ & 1 & $e^{j\frac{2\pi}{3}}$ & $e^{-j\frac{2\pi}{3}}$ & $e^{-j\frac{2\pi}{3}}$ \\ 
         & 0110 & 1 & $e^{j\frac{2\pi}{3}}$ & $e^{j\frac{\pi}{3}}$ & 0 & $e^{-j\frac{\pi}{3}}$ & 0 \\ \hline
    \end{tabular}
\end{table*}

\subsection{Gradient Descent Algorithm}
We first consider the ZCM criterion, proposed in \cite{ref6}, to unsupervisedly separate the colliding tags. Instead of observing this criterion through the resolution of a matrix pencil problem, we propose to achieve it through the minimization of an objective function using the gradient descent algorithm. The ZCM property can be written in compact form as
\begin{equation}
    s[n](|s[n]|^2 - 1) = 0,
\end{equation} 
for $1\leq n \leq N$. Thus, we consider the objective function as
\begin{equation}
\begin{aligned}
    J_0(\mathbf{w}) = & \, \frac{1}{N} \sum_{n=1}^N |\hat{s}[n]|^2 \left( |\hat{s}[n]|^2 - 1 \right)^2 \\
    = & \, \frac{1}{N} \sum_{n=1}^N  |\mathbf{w}^H \mathbf{x}[n]|^2 \left( |\mathbf{w}^H \mathbf{x}[n]|^2 - 1 \right) ^2.
\end{aligned}
\label{J0}
\end{equation}
The gradient descent algorithm underlying the minimization of this function is specified by the iterative process
\begin{equation}
    \mathbf{w}[l+1] = \mathbf{w}[l] - \mu \nabla J_0(\mathbf{w}) \big|_{\mathbf{w}= \mathbf{w}[l]},
\end{equation}
where $\mu$ is the step size, $l$ is the iteration number, and
\begin{equation}
    \nabla J_0(\mathbf{w}) = \frac{1}{N} \sum_{n=1}^N c[n]^* \mathbf{x}[n], 
\end{equation}
with
\begin{equation}
    c[n] = 2(|\mathbf{w}^H \mathbf{x}[n]|^2 - 1)(2|\mathbf{w}^H \mathbf{x}[n]|^2 + 1)(\mathbf{w}^H \mathbf{x}[n]). 
\end{equation}
We now propose the new correction metric corresponding to the new criterion (\ref{new_crit}) as
\begin{align}
J_1(\mathbf{w}) &= \frac{1}{N-2} \sum_{n=2}^{N-1} 
\left| \hat{s}[n-1] \hat{s}[n]^* \hat{s}[n+1] \right|^2 \notag \\
&= \frac{1}{N-2} \sum_{n=2}^{N-1} 
\left| \mathbf{w}^H \mathbf{x}[n-1] \right|^2 \notag \\
&\quad \times \left| \mathbf{w}^H \mathbf{x}[n] \right|^2 
\left| \mathbf{w}^H \mathbf{x}[n+1] \right|^2.
\end{align}

In order to raise the ambiguities encountered by the ZCM criterion in distinguishing the true beamformers from the spurious ones, we propose a new hybrid objective function to minimize, using the gradient descent algorithm, as
\begin{equation}
J_{01}(\mathbf{w}) = \lambda J_0(\mathbf{w}) + (1-\lambda) J_1(\mathbf{w}), 
\end{equation}
where the weighting parameter \( \lambda \in (0,1) \) is used to achieve a compromise between the ZCM metric, \( J_0(\mathbf{w}) \), and the new correction metric, \( J_1(\mathbf{w}) \).
As in the case of the ZCM metric taken alone, the minimization of the hybrid objective function \( J_{01}(\mathbf{w}) \) can be performed via gradient descent by updating the beamformer vector, \( \mathbf{w} \), as
\begin{equation}
\mathbf{w}[l+1] = \mathbf{w}[l] - \mu \nabla J_{01}(\mathbf{w}) \Big|_{\mathbf{w}=\mathbf{w}[l]}, 
\end{equation}
where
\begin{equation}
\nabla J_{01}(\mathbf{w}) = \lambda \nabla J_0(\mathbf{w}) + (1-\lambda) \nabla J_1(\mathbf{w}), 
\end{equation}
with
\begin{equation}
\begin{split}
\nabla J_1(\mathbf{w}) &= \frac{1}{N-2} \sum_{n=2}^{N-1} \big( 
c_- [n]^* \mathbf{x}[n-1] + c_0 [n]^* \mathbf{x}[n] \\
&\quad + c_+ [n]^* \mathbf{x}[n+1] \big), 
\end{split} 
\end{equation}

\begin{equation}
c_- [n] = 2 \left| \mathbf{w}^H \mathbf{x}[n] \right|^2 \left| \mathbf{w}^H \mathbf{x}[n+1] \right|^2 \left( \mathbf{w}^H \mathbf{x}[n-1] \right), 
\end{equation}
\begin{equation}
c_0 [n] = 2 \left| \mathbf{w}^H \mathbf{x}[n-1] \right|^2 \left| \mathbf{w}^H \mathbf{x}[n+1] \right|^2 \left( \mathbf{w}^H \mathbf{x}[n] \right), 
\end{equation}
and
\begin{equation}
c_+ [n] = 2 \left| \mathbf{w}^H \mathbf{x}[n-1] \right|^2 \left| \mathbf{w}^H \mathbf{x}[n] \right|^2 \left( \mathbf{w}^H \mathbf{x}[n+1] \right). 
\end{equation}
\section{Simulation Results}
\begin{figure}[tb]
    \centering
    \begin{subfigure}{0.49\columnwidth} 
        \centering
        \includegraphics[width=1\textwidth]{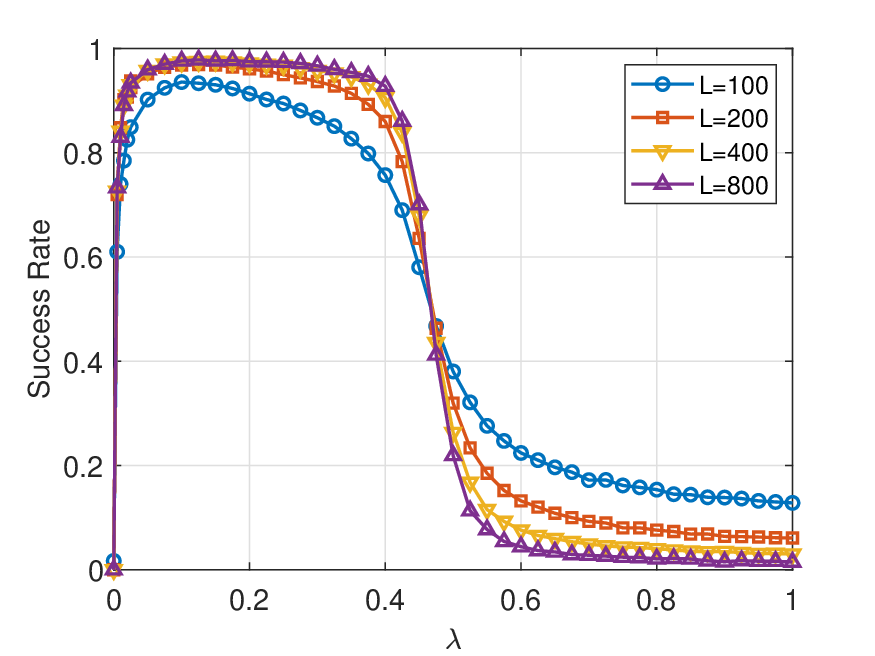}
        \captionsetup{labelformat=empty} 
        \caption{(a)}
        \label{fig1a}
    \end{subfigure}
    \hfill
    \begin{subfigure}{0.49\columnwidth}
        \centering
        \includegraphics[width=1\textwidth]{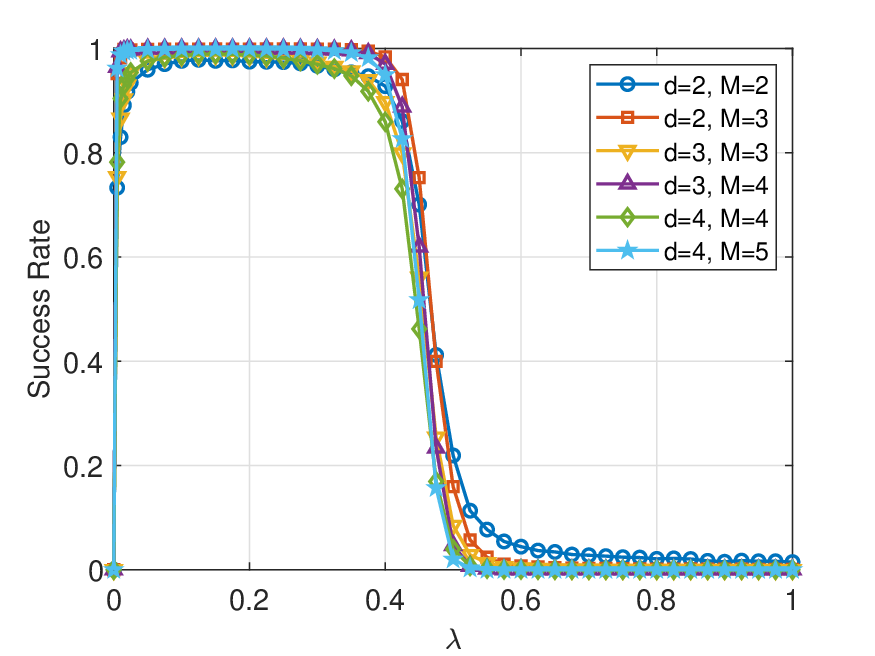}
        \captionsetup{labelformat=empty} 
        \caption{(b)}
        \label{fig1b}
    \end{subfigure}
    \hfill
    \begin{subfigure}{0.49\columnwidth}
        \centering
        \includegraphics[width=1\textwidth]{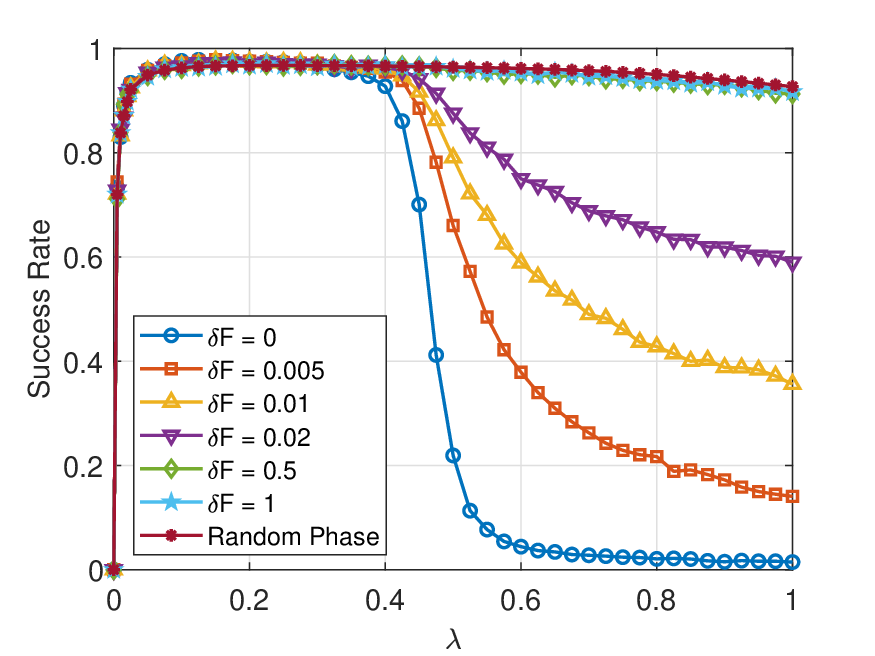}
        \captionsetup{labelformat=empty} 
        \caption{(c)}
        \label{fig1c}
    \end{subfigure}

    \caption{Success rate vs. the weighting coefficient $\lambda$ : (a) impact of the number of iterations $L$ when the number of antennas $M=2$ and the number of tags $d=2$, (b) impact of $d$ and $M$ when $L=800$, and (c) impact of random phase or frequency deviations when $M=d=2$ and $L=800$.}
    \label{fig1}
\end{figure}

We present simulation results to evaluate the performance of our proposed algorithm compared to the approach in \cite{ref6}. The packet size is set to \( N = 100 \). The number of iterations performed by the gradient descent algorithm is \( L \). The signal to noise ratio is \( \text{SNR}= \frac{E_b}{N_0}= 20 \, \text{dB} \), where \(E_b\) represents the energy per bit and \(N_0\) denotes the noise power spectral density. The step size is set to \( \mu = 10^{-2} \). A success is achieved if a tag message is identified. In Figs. \ref{fig1a} and \ref{fig1b}, perfect frequency synchronization is assumed. Fig. \ref{fig1a} shows the success rate versus the weighting coefficient \( \lambda \) for different iteration numbers \( L \), with \( d = 2 \) tags and \( M = 2 \) antennas. We notice that when the ZCM criterion is used alone (\( \lambda = 1 \)) as in \cite{ref6} or when \( \lambda \) is close to 1 (\( \lambda > 0.6 \)), the success rate is very low due to the ambiguities leading to spurious beamformers. When the new criterion is used alone (\( \lambda = 0 \)), the success rate is null. However, even a slight increase in $\lambda$ leads to a significant improvement in the success rate. Moreover, we remark that the range of $\lambda$ values providing high success rate expands as the number of iterations increases. This shows the robustness of our approach, as it remains effective over a broad range of $\lambda$ values. Fig. \ref{fig1b} shows the success rate versus \( \lambda \) when \( L = 800 \), for other values of \( d \) and \( M \). We notice that our proposed algorithm is also efficient for these configurations when $M=d$ for \( 0.025 < \lambda < 0.4 \) and even for lower values of $\lambda$ when $M>d$. Our results highlight the significant performance gap between our method and the approach in \cite{ref6}, demonstrating the superiority of our proposed solution. Fig. \ref{fig1c} shows the effect of frequency deviations and random phases on performance, for \( L = 800 \), \( d = 2 \) and \( M = 2 \). Frequency deviations are introduced in (\ref{symb}) as \( s_i[n] = b_i[n] \exp(2j\pi \delta F_in) \), for \( 1 \leq n \leq N \) and \( 1 \leq i \leq d \), where \( \delta F_i \) is a random variable uniformly distributed over \( [-\frac{\delta F}{2}, \frac{\delta F}{2} ]\). The curve “Random phase” corresponds to transmitted symbols \( s_i[n] = b_i[n] \exp(j\phi_i[n]) \) where \( \phi_i[n] \) is a random variable uniformly distributed over \( [0, 2\pi] \). We notice that as \( \delta F \) increases, the success rate improves since frequency deviations assist the algorithm in separating the tag messages when only the ZCM criterion is used or when \( \lambda \) is close to 1. Obviously, the best performance is achieved when random phases are considered. Importantly, Fig. 1c confirms that our algorithm remains effective whether or not phase and frequency drift imperfections are present, unlike the approach in \cite{ref6}, which performs poorly in the absence of such imperfections

\section{Conclusion}

In this letter, we considered an RFID system equipped with an antenna array at the reader. We banked on the ZCM criterion to unsupervisedly separate the colliding tags and showed that when ZCM is taken alone it could suffer from ambiguities leading to spurious beamformers. We presented a solution based on the gradient descent algorithm in order to minimize a hybrid objective function mixing the ZCM criterion and a new ambiguity-raising criterion. Simulation results shed light on the incapacity of the ZCM criterion to separate colliding tag signals when perfect frequency synchronization is achieved in the system. They also proved the effectiveness of the hybrid objective function in overcoming this issue.

\end{document}